\documentclass[11pt, oneside]{article}   	
\usepackage[margin=1.0in]{geometry}                		
\usepackage{graphicx}				
\usepackage{amssymb}

\newcommand{\lyaf}{Ly$\alpha$ forest}
\newcommand{\lya}{{Lyman$-\alpha$}}

\newcommand{\apj}{{Astrophys.~J.}}

\newcommand{\jcap}{{J.~Cosmology~Astropart.~Phys.}}
\newcommand{\mnras}{{Mon.~Not.~R.~Astron.~Soc.}}
\newcommand{\prd}{{Phys.~Rev.~D}}
\newcommand{\prl}{{Phys.~Rev.~Lett.}}

\usepackage[square]{natbib}             
\renewcommand{\cite}[1]{\citep{#1}}
	
\title{The DESI Experiment, a whitepaper for Snowmass 2013}
\author{M. Levi, C. Bebek, T. Beers, R. Blum, R. Cahn, D. Eisenstein, B. Flaugher, \cr K. Honscheid, R. Kron, O. Lahav, P. McDonald, N. Roe, and D. Schlegel,\cr 
representing the DESI collaboration}
\date{July 2013}							

\begin{document}

\maketitle

\begin{abstract}

The Dark Energy Spectroscopic Instrument (DESI) is a massively multiplexed
fiber-fed spectrograph that will make the next major advance in dark energy
in the timeframe 2018--2022.  On the Mayall telescope,
DESI will obtain spectra and redshifts for at least 18 million
emission-line galaxies, 4 million luminous red galaxies
and 3 million quasi-stellar objects, in order to: probe the effects of
dark energy on the expansion history using baryon acoustic oscillations
(BAO), measure the gravitational growth history through
redshift-space distortions, measure the sum of neutrino masses, and investigate the signatures of primordial inflation.  The resulting
3-D galaxy maps at $z<2$ and Lyman-alpha forest at $z>2$ will make
1\%-level measurements of the distance scale in 35
redshift bins, thus providing unprecedented constraints on cosmological
models.
\end{abstract}

\cleardoublepage

\section{Introduction}

DESI, the Dark Energy Spectroscopic Instrument, will be an exceptionally powerful facility for the Cosmic Frontier research program. DESI is a multi-fiber spectroscopic instrument that will be installed on the Mayall 4-m telescope to enable massively parallel measurements of galaxy redshifts.  The resulting 3D map of the Universe will enable baryon acoustic oscillation (BAO) measurements to chart the expansion history of the Universe, as well as red-shift space distortion measurements to chart the growth of structure.  DESI fits perfectly within the established Dark Energy program;  as a Stage-IV BAO experiment, it complements the DES and LSST imaging surveys, whose strengths are weak lensing and Type Ia supernovae. With DESI, the US can maintain a healthy program cosmology and fill the hiatus between the end of DES and the start of LSST.  CD-0 (mission need) was awarded to DESI by DOE in September 2012, CD-1 is scheduled for January 2014, and first light is planned for 2018.

The large-scale structure of the Universe provides a key test for cosmological models. The fundamental measurement of large-scale structure is the correlation function that measures the increased (or decreased) likelihood of finding galaxies separated by some distance relative to that in a purely random distribution.  DESI observations of this correlation function will allow us to probe diverse aspects of cosmology, from the nature of dark energy, to the neutrino mass hierarchy and absolute mass scale, to signatures of inflation. 

Acoustic waves initiated by small overdensities in the very early Universe left an imprint on the matter distribution when the plasma turned into neutral atoms.  The baryon acoustic oscillation (BAO) technique measures this imprinted pattern as a function of the redshift, that is as a function of the time from the Big Bang.  The power and promise of the BAO technique on which DESI is based have become evident in recent years.  Of the four major techniques to study dark energy (supernovae, clusters, weak lensing, and BAO), progress in BAO has been especially dramatic from the time of its first experimental observation in 2005.  The success of the BOSS survey has demonstrated the robustness of the BAO technique and its relative immunity to systematic errors, and has demonstrated that ÒreconstructionÓ can be used to reduce the influence of non-linear effects.  BOSS has also detected the BAO signal in Lyman-alpha forest data in quasar spectra, extending the BAO method to higher redshifts with the first measurement of the expansion history of the Universe at redshift $z>2$, and demonstrated the feasibility of other new techniques such as redshift space distortions (RSD), which probe the growth of structure as well as spacetime geometry, and the Alcock-Paczynski (AP) effect, which provides a complementary measure of cosmic geometry. Other measurements include constraints on the sum of neutrino masses, and on primordial non-Gaussianity from the early Universe. A massive spectroscopic survey thus provides a rich data set for multiple constraints on cosmology.

In this white paper  we first review the dark energy program and state of the field for BAO surveys. Then we provide a summary of the DESI design, the parameters of the planned reference mission, and its scientific reach. 

\section{The Dark Energy Program }
Prior to 1998, it was assumed that the expansion of the universe would slow down due to the gravitational attraction of matter.  However, in 1998, two teams using ground-based and space-based supernova measurements found that the expansion of the universe is actually accelerating.  The previously unsuspected energy in the universe causing this acceleration has been dubbed "dark energy."  It has now been established that only about 5\% of the universe is made of the normal, visible matter, including the earth, stars and all other objects we see.  The remaining 95\% of the universe consists of dark matter and dark energy whose fundamental natures remain mysterious.

To date, there are no compelling theoretical explanations for the dark energy, and observational exploration is the focus of the effort. Understanding the nature of dark energy will provide exciting new discoveries that can change the way we view the universe and could have profound implications for fundamental physics.

The U.S. is currently the leader in the exploration of dark energy, which will offer new insights and a deeper understanding of fundamental physics and the makeup and ultimate fate of the universe.  Planning for future dark energy experiments commenced in 2006 with Dark Energy Task Force  (DETF) report, which was approved by HEPAP.\footnote{The Dark Energy Task Force was a joint subpanel of the High Energy Physics Advisory Panel and the Astronomy and Astrophysics Advisory Committee charged with providing advice to the DOE, NSF, and NASA on the optimal intermediate- and long-term programs to study dark energy.\cite{DETF}}  The DETF created a figure of merit (FOM) to describe the improvement in precision on the measured values of the parameters describing the expansion of the Universe and its time evolution. In the DETF report, dark energy experiments were classified in stages with higher numbered stages reflecting more sophisticated and powerful experiments with higher FOMs. Stage-I represented what was then known in 2006.  Stage-II represented the anticipated state of knowledge upon completion of the then-ongoing dark energy projects. Stage-II is now complete for SN and BAO experiments.  Stage-III comprised near-term, projects (eg. BOSS and DES) that would increase the DETF FOM by a factor of 3 relative to Stage-II.  Stage-IV comprised future proposals to increase the DETF FOM by a factor of 10 over Stage-II. 

The recent Dark Energy Science Plan Task Force  (aka, Rocky-III) has reviewed the state of the field since the 2006 DETF report and the recommendations have been updated along with prioritizing the next steps to be taken in this research area.\cite{Albrecht2012}  At the time of the 2006 DETF report, BAO was a relatively new technique. Subsequent experiments, such as BOSS, with DOE support, have shown that this technique has a unique role in extending our knowledge of dark energy.  Rocky-III highlighted a future wide-field spectroscopic survey exploiting the BAO technique as the highest priority next experiment to be undertaken by the dark energy community.  DESI fulfills these recommendations as a Stage-IV spectroscopic dark energy experiment and complements LSST, the planned Stage-IV dark energy imaging survey.

\section{Baryon Acoustic Oscillations:  State of the field}
DESI will be the first Stage-IV dark energy experiment, surveying 20--30M luminous red galaxies (LRGs), emission line galaxies (ELGs) and quasars (QSOs). In the time period between now and the planned start of DESI in 2018, BOSS will finish its survey of 1.5M galaxies, eBOSS may eventually double that number, and HETDEX will survey approximately one million Lyman-alpha emitters at high redshift.  These experiments will develop further the BAO technique, providing an even stronger basis for DESI, whose scope and measured volume will far exceed its predecessors. Concurrent with DESI, the Sumire/PFS experiment at the Subaru telescope will survey 4M galaxies to higher redshift than DESI, while the proposed 4MOST at ESO/Vista and WEAVE at WHT will have limited dark energy programs; their focus is on the follow-up of GAIA and eRosita objects.  

\begin{figure}
\begin{center}
\includegraphics[width=5truein]{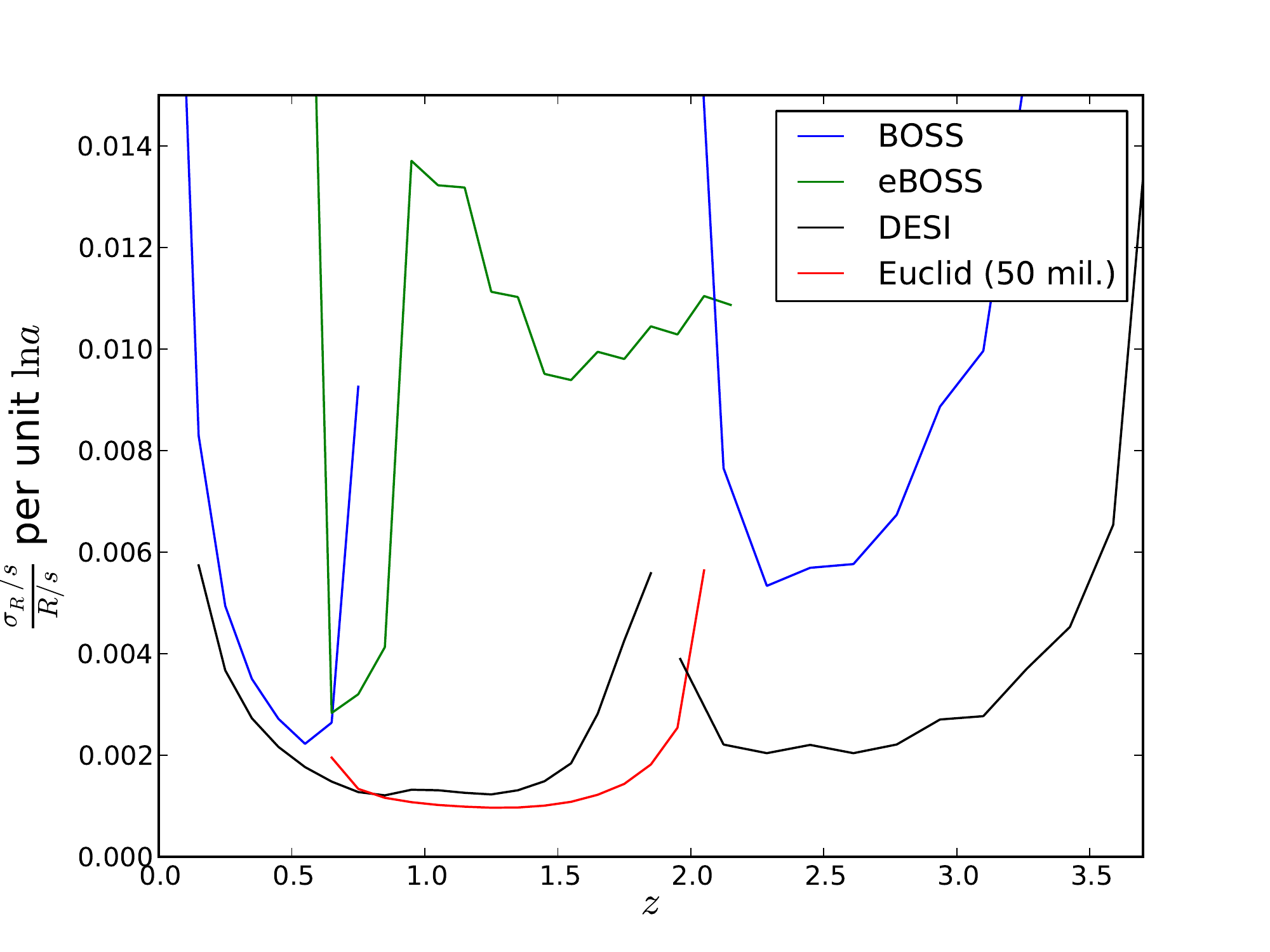}

\caption{ 
\small
The power of DESI is in both the precision and the wide range of redshifts it will cover, making it competitive even with the Euclid space-based mission.
Shown are the fractional error on the 
BAO distance scale (isotropic dilation factor), as a function of redshift, {\it per unit $ln(a)$}
(in other words, the effect of any arbitrary redshift bin width $\Delta z$ is 
removed in this plot).  
Errors from the \lyaf\ measurement, which dominate at $z>1.8$, are 
computed following McDonald \& Eisenstein (2007), with a modest but significant
additional contribution from cross-correlations with quasar density. 
We assume here an optimistic 50 million galaxies for EUCLID.}\label{Rfigure}

\end{center}
\end{figure}

The European EUCLID satellite will launch sometime early in the next decade.  It is also a Stage-IV dark energy experiment. EUCLID is designed to measure weak lensing and BAO in the redshift range 1.4 to 2.0. The NASA/WFIRST mission is in formulation.  In its current design (using a 2.4-meter mirror, intended for 2022 launch), the BAO survey would cover a much smaller area than DESI but at much higher sampling density, making the two experiments complementary.  Finally, new experimental techniques, such as BAO using observations of the 21cm line of neutral hydrogen hold significant future promise, but their designs do not allow the more revealing broadband power spectrum measurements of DESI. Such experiments include CHIME and BAOBAB, both of which have pathfinder instruments under construction.  These surveys are summarized in Table~\ref{tableexpts} and the science reach for the BAO surveys is illustrated in Figure~\ref{Rfigure}.  The power of DESI is in both the precision and the wide range of redshifts it will cover, making it competitive even with the Euclid space-based mission.

\begin{table}
\caption{Summary of current or planned BAO capable spectroscopic surveys.
\cite{Eisenstein2001}\cite{Hogg2005}
\cite{Drinkwater2010}\cite{Scrimgeour2012}
\cite{Eisenstein2011}\cite{Bolton2012}
\cite{Hill2008}
\cite{Abdalla2012}   \cite{Schlegel2011}
\cite{Ellis2012}
\cite{deJong2012}
\cite{Amiaux2012}
}
\label{tableexpts}
\small
\begin{center}
\begin{tabular}{lccccc}
\hline
\hline
Instrument &	Telescope 	& Nights/ year &	No. Galaxies &sq deg &	Ops Start 
\\
\hline
SDSS I+II	&APO 2.5m		&dedicated	&85K LRG	&7600	&2000\\
Wiggle-Z	&AAT 3.9m		&60	&239K	&1000	&2007\\
BOSS	&APO 2.5m		&dedicated	&1.4M LRG+160K Ly-$\alpha$	&10000	&2009\\
HETDEX	&HET 9.2m 		&60	&1M	 &420	&2014\\
eBOSS	&APO 2.5m		&180	&600K LRG + 70K Ly-$\alpha$	&7000	&2014\\
DESI	&NOAO 4m		&dedicated	&+20M + 800k Ly-$\alpha$ & 14000	 &2018\\
SUMIRE PFS 	&Subaru 8.2m 		&20	&4M	 & 1400	& 2018\\
4MOST	&VISTA 4.1m		&shared facility	&6-20M bright objects	&15000	&2019\\
EUCLID	&1.2m space		&dedicated	&52M	&14700	&2021\\
\hline
\hline
\end{tabular}
\end{center}
\end{table}

\section{DESI Instrument Reference Design}
The design of the DESI instrument is set by the key science project and operational requirements, primary ones being:

\begin{itemize}
\item Survey operates from 2018 through 2022
\item 14,000 -- 18,000 sq. deg. BAO/RSD redshift survey
\item Targets are LRGs, ELGs and QSOs including Ly-$\alpha$ forest
\item $20 - 30$ million targeted galaxies and QSOs for $0.5 < z < 3.5$
\item Spectroscopic resolution sufficient for redshift error $< 0.001(1+z)$ 

\end{itemize}

\begin{figure}
\begin{center}

\includegraphics[width=6truein]{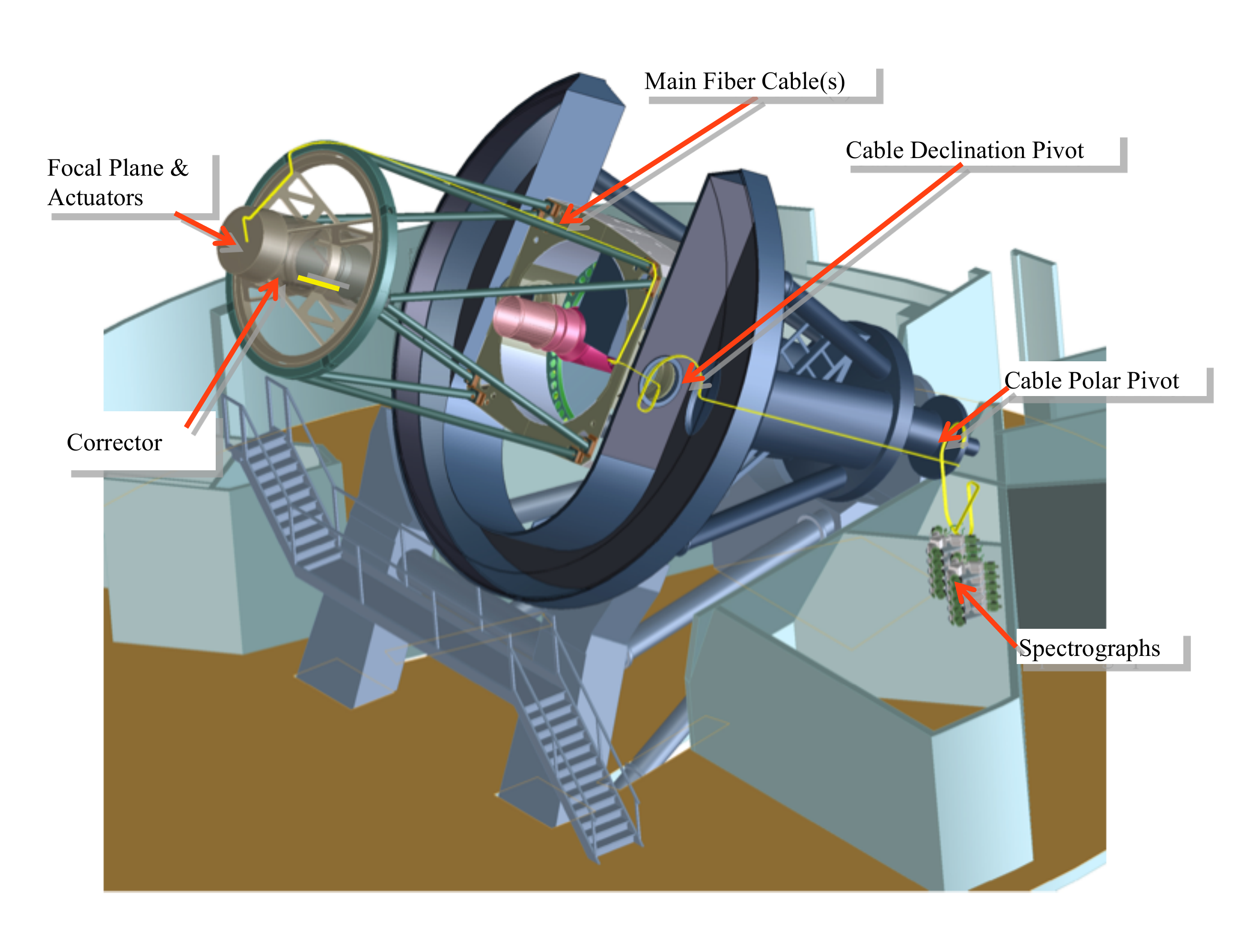}
\caption{
\small
The Dark Energy Spectroscopic Instrument is shown in the figure.  A new corrector delivering a 8 deg$^2$ field-of-view is installed at the Mayall 4m telescope and brings light to an array of 5000 fibers.  Each fiber tip can be moved independently to a random galaxy position by mechanical actuators under computer control.  The fibers are brought to a system of spectrographs that can simultaneously observe the dispersed light from all 5000 fibers.}\label{instrument}

\end{center}
\end{figure}

Performing a wide, deep spectroscopic survey of a large volume of the Universe with a density $>1500$ galaxies/deg$^2$ in a five-year survey requires a high throughput spectrograph capable of observing thousands of spectra simultaneously. The DESI project is designed to achieve these ambitious goals.  The instrument components are 1) prime focus corrector optics to achieve a wide field of view, 2) focal plane with robotic fiber positioners, 3) fiber optics cable management system, 4) spectrographs, 5) a real-time control and data acquisition systems, and 6) a data processing pipeline that ingests raw data from the detectors and produces calibrated spectra useful for cosmological investigation.  A conceptual drawing of the telescope and instrument is shown in Figure~\ref{instrument}.   

Here we briefly discuss critical instrument parameters and their connection to the key science projects and operational constraints. The Mayall telescope has been identified as the telescope most suitable for DESI, and DOE and NSF are currently discussing the terms for its use.  The Mayall is a 4-m telescope capable of supporting complex prime focus instruments and attaining a field of view from $2.5-3.0$ degrees diameter. Combining the field of view, survey duration, galaxy spectra count, spectral resolution, and signal-to-noise leads to a focal plane design that accommodates $4000 - 5000$ repositionable optical fibers on a 12 mm or smaller pitch. The entire focal plane needs to be reconfigured for the next exposure in less than a minute with the fiber tips placed with an accuracy of $5 \mu m$ r.m.s. with robotic positioners.  The repositioning time overlaps telescope slew and readout of the spectrograph detectors.  Also on the focal plane are a star tracker system for telescope guiding and sensors for corrector-focus determination.  Focus information derived from these sensors drives a six-axis hexapod to adjust the optical corrector barrel and the focal plane position.
 
A system of optical fibers transports the light from the focal plane to the spectrographs.  The fibers are collected from the focal plane in take-up spool boxes and then bundled into cables for the approximately 40-meter run to the spectrographs.  Along their route are support mechanisms to handle telescope polar and azimuth motions. The fibers terminate in the spectrograph room, first in take-up spools and then slit arrays, each feeding a spectrograph.

A 500-fiber slit array feeds the spectrograph Schmidt collimator followed by cameras.  The full spectral bandpass is 360 to 980 nm divided between three arms, using dichroics.  The blue spectral resolution of $R>1500$ is derived from Lyman-alpha forest observation requirements from $2.2 < z < 3.5$.  The red arm resolution of $R>3000$ is derived from the LRG 400 nm break from $0.5 < z < 1.0$.  The NIR arm resolution of $R>4000$ is required to resolve ELG [OII] doublet from $0.5 < z < 1.7$.  Each camera has 4kx4k, 15 $\mu$m pixel CCDs with individual control and digitization electronics. The baseline design achieves an end-to-end throughput of greater than 40\% (excluding the telescope or atmosphere).

The DESI exposure sequence is orchestrated by the instrument readout and control system. In order to maximize survey throughput, many operations are interleaved and the dataflow system uses a pipelined architecture.  The focal plane and the fiber positioner are set up for the next exposure while the previous image is being digitized and readout. At the same time the telescope slews to a new field. At a typical pixel clock rate of 100 kHz CCD readout will take approximately 42 seconds. In order to achieve the required survey efficiency the time between exposures has to be less than 60 seconds.

\section{DESI Reference Mission}

\subsection{Survey Footprint}
\begin{figure}
\begin{center}

\includegraphics[width=5truein]{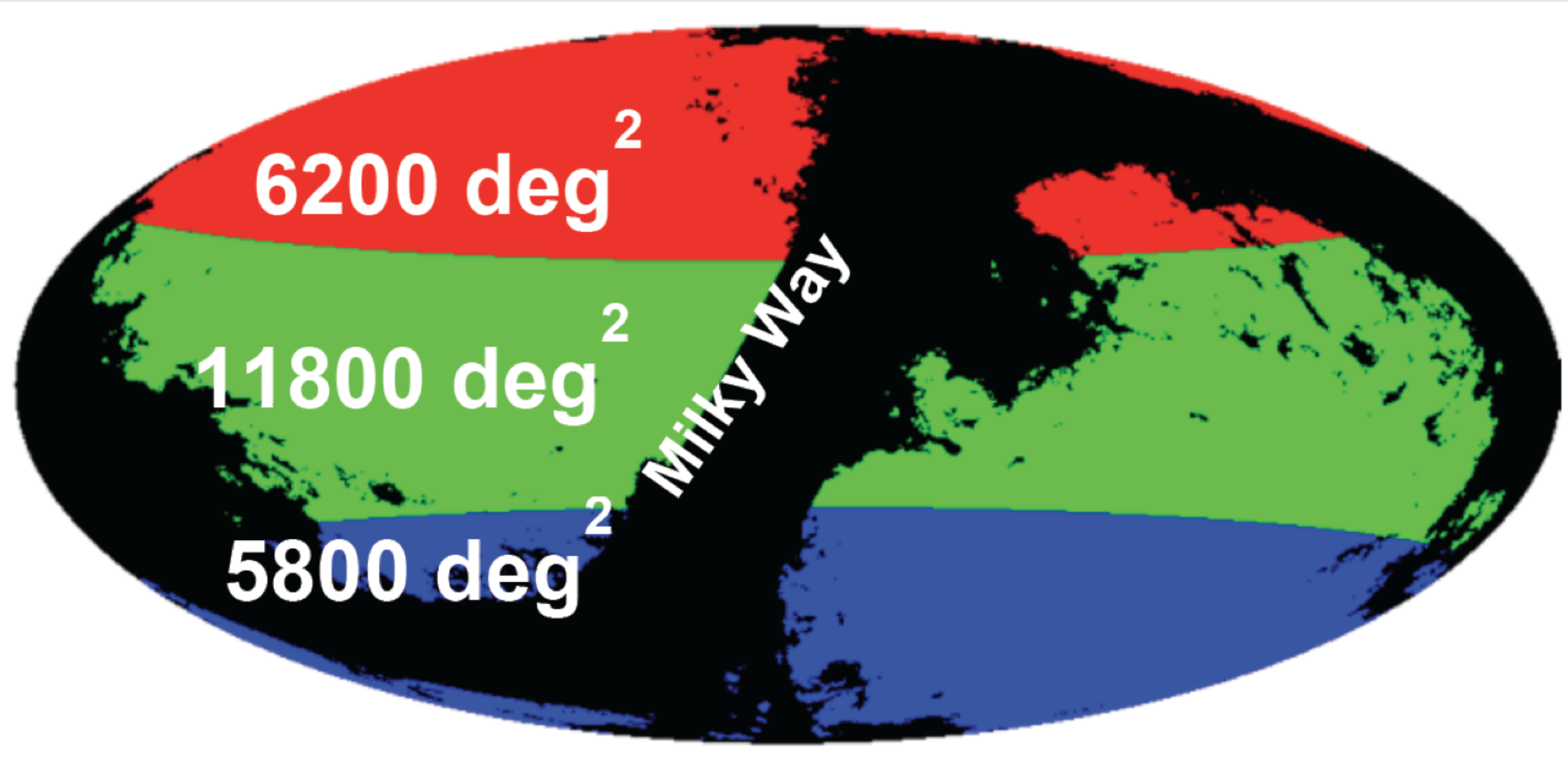}
\caption{
\small
Aitoff projection of the low-reddening ($E(B-V) < 0.094$) extragalactic footprint suitable for the DESI survey.  The total footprint is 23,800 deg$^2$, where 11,800 deg$^2$ is the equatorial region.  In this calculation the northern + equatorial sky visible from KPNO totals 18,000 deg$^2$. The imaginary lines are at Dec$=-28$ deg and Dec$=+28$ deg.}\label{Aitoff}

\end{center}
\end{figure}

The full-sky extragalactic footprint appropriate for optical BAO experiments is approximately 24,000 deg$^{2}$.  This is defined as the region of low extinction from the Milky Way. An equatorial zone ($-20 < Dec < +20$ deg) is accessible from all telescope resources for DESI target selection.  This defines a minimum 8,800 deg$^{2}$ extra-galactic footprint viewable from all telescopes for target selection and DESI spectroscopy.  Extending these cuts to $-28 < Dec < +28$ deg could enlarge this equatorial footprint to 11,800 $deg^{2}$ (Figure~\ref{Aitoff}).  DESI sited at Kitt Peak can access an additional 6,200 deg$^{2}$ of Northern sky.  These regions are shown pictorially in Figure~\ref{Aitoff}.   The sky area for the DESI reference mission is 14,000 deg$^{2}$, with 18,000 deg$^{2}$ as a stretch goal.

\subsection{Target Classes and Target Selection}
 
DESI will use four target classes extending from redshift $z>0.5$ to as high as $z<4.0$.
 
\begin{enumerate}
\item Luminous Red Galaxies (LRGs) are the reddest, most luminous galaxies in the Universe at redshifts $z < 1$.  These galaxies can be efficiently identified in imaging surveys in combination with the completed Wide-Field Infrared Survey Explorer (WISE) satellite data.  The minimum required supplement to the WISE data is a single optical band in r, i, or z-band to a depth of $\approx 23.5$ mag.  The addition of a second optical band improves the sculpting of the redshift distribution.
\item Emission Line Galaxies (ELGs) are star-forming galaxies with strong emission lines and characteristic OII and OIII doublets at a rest frame 327 nm and 500 nm, which make them suitable for the DESI redshift survey.   These can be selected at the number densities for a BAO survey with two filters, g-band in combination with either r-band or i-band.  The addition of u-band or z-band would improve the selection of those objects with bright emission lines and permit better sculpting of the redshift distribution.
\item QSOs are an excellent tracer population at redshifts $1 < z < 2$ where their numbers peak, although at number densities less than those of ELGs.  They can be selected from g,r,i-band photometry with a high contamination rate ($\approx 50\%$) from stars.  The addition of deep u-band data, multi-epoch photometry in any optical band, or existing WISE data reduces this contamination rate.
\item Lyman-alpha QSOs at $z > 2.1$ represent the only opportunity for DESI to measure dark energy at these high redshifts.  The selection methods are the same as those for the QSO tracers.

\end{enumerate}

\subsection{Imaging Requirements}
The ensemble of DESI target classes requires pre-imaging survey in two optical bands in addition to the existing (full-sky) WISE satellite data.  The depths of these bands is $\approx 23.5$ mag (5$\sigma$ AB), which is about one magnitude deeper than SDSS imaging, but significantly shallower than the weak lensing surveys of SNLS, DES and LSST.  The addition of a third optical band would further improve the ELG target selection. Imaging of the Northern footprint will be acquired through one or more of the following instruments: 19 deg$^2$/night with 4-m Mayall MOSAIC, 34 deg$^2$/night with 3.5-m WIYN pODI, or 65 deg$^2$/night with the Zwicky Transient Factory at Palomar. Discussions are also underway for access to the 4-m CFHT Megacam and its planned upgrade. For the equatorial region, imaging data will be obtained at 160 deg$^2$/night with DECam.  The suitability of Pan-STARRS1 survey data is also under study.  The collaboration is actively engaged in activities that will secure the required imaging resources from one or more of these instruments/surveys.

\section{DESI Scientific Performance}

Here we focus on the DESI key science theme, namely the physics of dark energy.  Numerous community reports have identified two principal cosmological measurements for the study of dark energy:  1) the expansion history of the Universe and 2) the growth rate of large-scale structure.  DESI observations will study both.  In addition, we describe how DESI will provide an important constraint on the sum of neutrino masses, one that is complementary to other approaches that can determine the mass hierarchy or establish the neutrino mass scale through precision measurements of rare processes involving neutrinos.  We provide the DETF FOM calculations for DESI at the end of this section.

\subsection{Baryon Acoustic Oscillations}
A central goal of DESI is the measurement of the expansion history of the Universe by the baryon acoustic oscillations (BAO) imprinted in the clustering of galaxies, quasars, and the intergalactic medium. The BAO is the most robust way to extract cosmological distance information from the clustering of matter and galaxies.  It relies only on very large-scale structure and it does so in a manner that is differential in redshift, allowing us to separate the acoustic peak from uncertainties in galaxy bias or most systematic errors in the data.  Both analytic and numerical studies indicate that the acoustic peak is a standard ruler to better than 1\% in a wide range of Cold Dark Matter (CDM) cosmologies, and indeed it is thought it can be calibrated to better than 0.1\%.  The method is likely to be limited by statistics rather than systematics all the way to the full-sky limit at $z<3$.\cite{Weinberg2012}

The challenge of the BAO is to survey enormous volumes of the Universe with an acceptable number density of tracers. The ÒRocky-IIIÓ dark energy community panel report identified spectroscopic surveys as the most effective way to address this question.  The required tracer density is approximately $\approx 1500$ per deg$^{2}$, a small fraction ($\approx 3\%$) of the abundance of galaxies.  This means that one can choose targets for observational convenience, e.g., the brighter or more efficiently measurable targets at a given redshift.  

In the context of realistic multi-object galaxy surveys, optimization of BAO performance for dark energy always drives one to maximize sky area, i.e., go wide, then deep.  The gains from observing the more accessible (brighter or lower redshift) galaxies over more area outweigh the gains from observing the more expensive to access fainter, higher redshift galaxies. With the strong scientific incentive for DESI to cover as much sky as possible, at least 14,000 square degrees, and preferably the full 18,000 square degrees, of low-extinction sky should be observed from a mid-latitude site.  

\subsection{Redshift-space Distortions} 
In addition to the BAO feature, large-scale structure contains additional information about dark energy through redshift-space distortions (RSD), which probe the growth rate of structure with precision that is comparable to that from weak lensing.  This provides information on the potential metric perturbations away from GR.  Large-scale structure grows in amplitude because matter flows in the direction of gravitational forces.  The redshifts of galaxies traveling in these flows include the line-of-sight component of the velocity field.  By comparing the apparent clustering of pairs of galaxies along the line-of-sight to pairs transverse on the sky, we can directly infer the flow of matter and hence measure the growth rate of structure.  This is known as the redshift-space distortion (RSD) method.  

One can also measure the growth rate of structure by comparing the amplitude of cosmic density perturbations at two different redshifts, for example by using a weak-lensing survey.  In simple models of structure formation, the in-fall measured by RSD is equivalent to this lensing measurement.  In more complicated models, RSD and lensing measure different components of the relativistic metric.  Since our interest in the growth of structure measurement is to test general relativity on cosmic scales, this test of the metric is of considerable interest.  
  
Like BAO, the RSD method requires a map of the large-scale structure of the Universe and the DESI redshift survey of galaxies will produce an exceptional measurement of the growth rate of structure via the RSD method.  

\subsection{Neutrino Properties}
\begin{figure}
\begin{center}

\includegraphics[width=5.5truein]{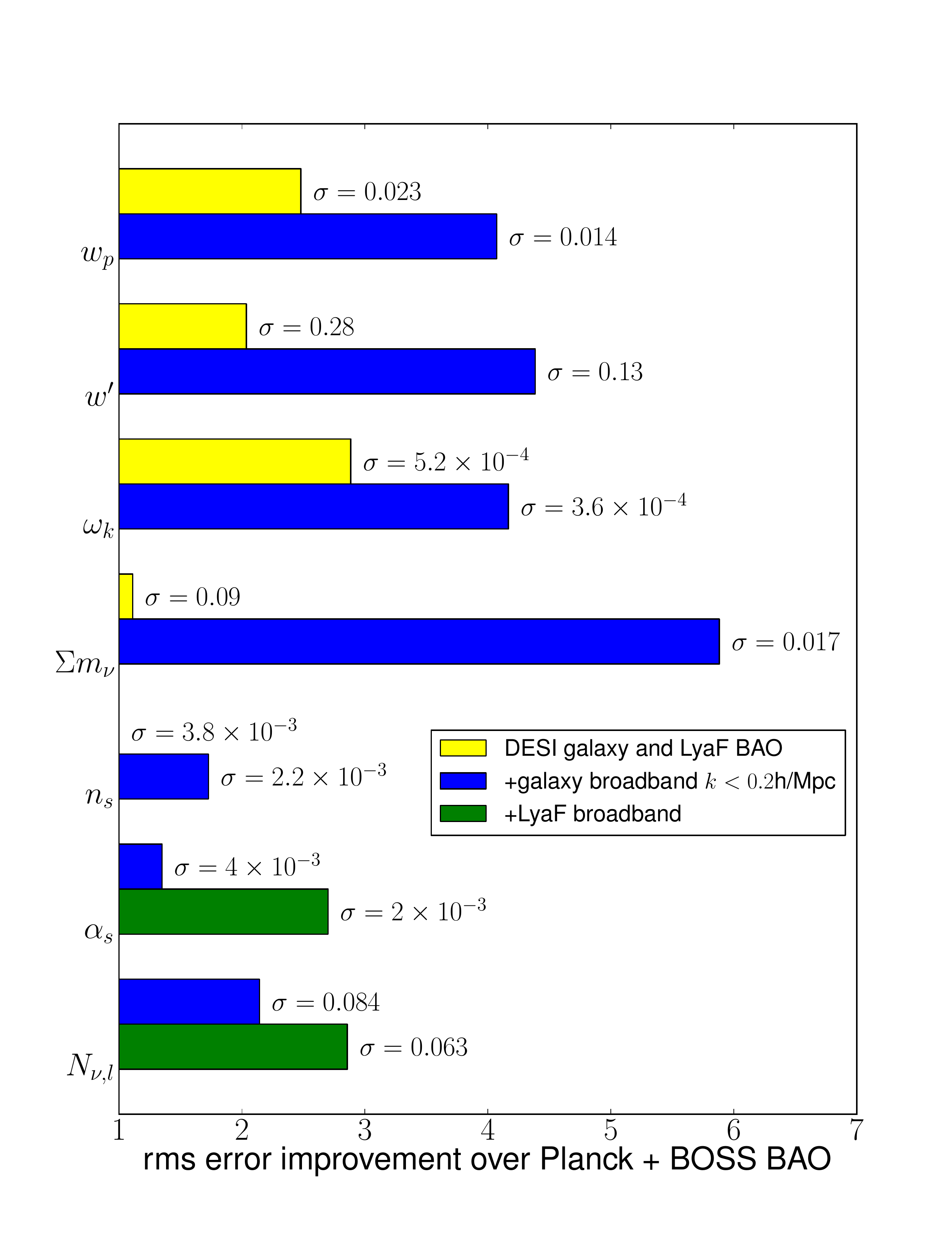}
\caption{
\small
Improvement in parameters over Planck plus BOSS BAO 
(projected finished {\it Planck},
including polarization). Yellow bars show the improvement with DESI BAO, including \lyaf. Blue
bars show the additional improvement when optimistic broadband power is
used, with $k<0.2$ h/Mpc. Green additionally adds potential constraints
from broadband (including small-scale 1D) \lyaf\ power.  Dark energy parameters are defined by the 
equation of state $w(z)=w_p + (a_p-a)w^\prime$, where $a_p$ is
chosen to make the errors on $w_p$ and $w^\prime$ independent. 
$\omega_k \equiv \Omega_k h^2$ is the equivalent physical density of 
curvature. $\sum m_\nu$ is the sum of masses of neutrinos, in eV. 
$\alpha_s$ and $n_s$ parameterize the inflationary perturbation
power spectrum, i.e.,  $P_{\rm inflation}(k) 
\propto k^{n_s +\frac{1}{2}\alpha_s \ln\left(k/k_\star\right)}$. }\label{DESI_reach}

\end{center}
\end{figure}

In addition to the measurement of the expansion history of the Universe and the growth rate of large-scale structure, DESI has an enormously broad scientific reach including tests of primordial inflation (see Figure~\ref{DESI_reach}).  DESI will also provide an important constraint on the sum of neutrino masses, one that is complementary to other approaches that can determine the mass hierarchy or establish the neutrino mass scale.  The power spectrum derived from the DESI redshift maps is uniquely suited for measuring the absolute neutrino masses and the effective number of neutrino species.  The mass measure will inform the neutrino mass hierarchy question, with a significant result if the neutrino masses are near the minimum consistent with known mass-squared splittings.\footnote{This subsection on neutrino properties is an updated version of material from arXiv:1106.1706.}

The effects of neutrinos in cosmology are well understood. They decouple from the cosmic plasma when the temperature of the Universe is about 1 MeV, just before electron-positron annihilation. While ultra-relativistic, they behave as extra ÒdarkÓ radiation with a temperature equal to (4/11)1/3 of the temperature of the cosmic microwave background. As the universe expands and cools, they become non-relativistic and ultimately behave as additional dark matter.

Neutrinos have two important effects in the early universe. First, as an additional radiation component they affect the timing of the epoch of matter-radiation equality. Second, the process of neutrinos becoming non-relativistic imprints a characteristic scale in the power spectra of fluctuations. This is termed the Ôfree-streaming scaleÕ and is roughly equal to the distance a typical neutrino has traveled while it is relativistic. Fluctuations on smaller scales are suppressed by a non-negligible amount, of the order of a few percent. This allows us to put limits on the neutrino masses.

In principle, if multiple mass eigenstates are present, the cosmological measurements should be sensitive to individual mass states, but in practice they are strongly degenerate and the only quantity cosmology really measures is the sum of neutrino masses.  Projections for data sets similar to the DESI BAO survey have been carried out by Stril, Cahn, \& Linder (2010) and Thomas, Abdalla \& Lahav (2010).  Because one of the splittings of the squares of the neutrino mass is about $2.43 \times 10^{-3} eV^2$ (KamLAND Collaboration, 2005), we know that at least one neutrino has a mass of at least 0.05 eV. If neutrinos have an inverted mass hierarchy, the minimum sum of the neutrino masses is roughly twice this (since the other splitting is considerably smaller).  The $1\sigma$ limits obtained on the sum of neutrino masses should be 0.024 eV (integrating in k-space up to $k_{max} = 0.1$h/Mpc.  More stringent constraints (rms of 0.017 eV) are possible if non-linearities are controlled up to $k_{max} = 0.2$h/Mpc; this is an active area of theoretical study.
 
DESI will make a greater than 2-sigma detection of the sum of the neutrino masses, even in the case of a normal mass hierarchy, and it could rule out the inverted mass hierarchy at a similar confidence level. Either of these would be a major result with important repercussions for particle physics as well as cosmology.
 
The other parameter relevant for neutrino physics is the effective number of neutrino species $N_{\nu}$, which parameterizes the energy density attributed to any non-electromagnetically interacting ultrarelativistic species (including e.g. axions) in units of the equivalent of one neutrino species that fully decouples before electron-positron annihilation. The value for the standard cosmological model is $N_{\nu}$ = 3.04. The detection of any discrepancy from the expected value would be a major result, indicating either a new particle, non-standard neutrino physics or some other exotic new physics. The effective number of neutrino species will be measured to an accuracy much better than unity, providing strong constraints on the alternative models involving extra sterile neutrinos, axions or partly-thermalized species.

\subsection{Figure-of-Merit}
In this section we review the calculations of figures-of-merit for dark energy and for tests of General Relativity using redshift-space distortions and modeling of the power spectrum.  We have computed the scientific reach of DESI as captured by the DETF FOM, assuming a number of scenarios for survey area (see Table~\ref{tableFOM}).  A 14k sq. deg. survey would find $\sim24$M galaxies and give BAO+RSD FOM of $\sim205$, to be compared with BOSS at FOM $\sim22$, both with Planck priors.  While a 18k deg$^2$ would measure 32M galaxy redshifts and give a FOM of $\sim261$.  Alternatively, the DESI could trade some area for greater density and attain 40M galaxy redshifts over 14k deg$^2$ (indicated as 14k-HD).

\begin{table}
\caption{
\small
Dark energy performance, all fits include Planck.   Broadband power (indicated as +PS) for $k < 0.1 h/Mpc$.  DE FOM is per DETF.  MOG $FOM = \sigma_{\gamma} \times \sigma_{G_{9}}$.  Where WL (indicated as +WL) is included, assumes same sky area as BAO, but not overlapping. With broadband and weak lensing, DE FOM is marginalized over MOG parameters. The survey marked Ò14K-HDÓ is taken at a 67\% higher target density.  MOG parameters are taken from \cite{FoMSWG}. Ref: P.  McDonald.}
\label{tableFOM}
\small
\begin{center}
\begin{tabular}{lcccccc}
\hline
\hline
Survey Size &	No. of &	BAO &	DE & MOG &	DE & MOG\\
&	Galaxies	&	(galaxy+Ly-$\alpha$) & FOM & FOM & FOM & FOM\\
&	 		&	DE FOM & +PS & +PS & +WL & +WL\\
\hline
BOSS&		1.5M&	21&		22&		41&		&	 \\
DESI 5k&	8M&		55&		76&		728&	103&	1149\\
DESI 10k	&	17M&	104&	149&	1203&	190&	2042\\
DESI 14k	&	24M&	140&	205&	1497&	257&	2723\\
DESI 14k-HD&40M&	187&	276&	1816&	325&	3267\\
DESI 18k	&	32M&	174&	261&	1759&	322&	3369\\
\hline
\hline
\end{tabular}
\end{center}
\end{table}
 
DESI will use a combination of emission-line galaxies (ELGs), luminous red 
galaxies (LRGs), and quasars to trace large scale
structure out to redshift $z \sim 1.7$ and \lya\ forest absorption towards 
nearly one million high-redshift quasars to measure BAO
at redshifts $1.9 < z < 4$.  As detailed in the Table~\ref{tableDESIsimple} 
below, 
forecast BAO errors on the distance scale are $0.35 - 1.1\%$ per
{\it $\Delta z = 0.2$} redshift bin out to $z = 1.7$, with
an aggregate precision of 0.17\%.  Typical distance errors per bin from the 
\lya\ forest, augmented by cross-correlations with quasar density, 
are $\sim 1\%$, with an aggregate precision of 0.37\%.   
The unique power of \lya\ forest is shown in Figure~\ref{Rfigure}.  Structure growth is measured via redshift-space distortions in the galaxy and 
quasar distributions, with per-bin errors on
the parameter combination $f\sigma_8$ below 2\% over the range $0.2 < z < 1.6$ 
and aggregate precision of 0.35\%.

\begin{table}
\caption{
\small
Basic numbers for DESI. $\frac{\sigma_{R/s}}{R/s}$ is the projected percent 
error on a measurement of an isotropic dilation of the BAO distance scale.
$V$ is the volume covered by DESI in this redshift slice,
$\frac{dN_X}{dz d{\rm deg}^2}$ 
is the number of objects of type $X$ (LRGs, ELGs, or QSOs) per unit z per
square degree.
$\frac{\sigma_{f\sigma_8}}{f\sigma_8}$ represents the error on
the velocity divergence power spectrum aggregated over 
$k<0.2 h {\rm Mpc}^{-1}$, conventionally quoted as a percent error on 
$f(z) \sigma_8(z)$ where $f=d\ln D/d\ln a$ where $D$ is the linear growth 
factor and
$a$ the expansion factor, and $\sigma_8(z)$ is the rms amplitude of 
density perturbations at redshift $z$. 
Projections for BAO distance errors are made as described in 
Seo \& Eisenstein (2007).
Projections for RSD errors are made as described in McDonald \& Seljak (2009).
Biases are assumed to follow
$b(z)=b_0 \left(D(z=0)/D(z)\right)$, where $b_0=1.7$, 0.84, and 1.2 for 
LRGs, ELGs, and QSOs, respectively. 
\label{tableDESIsimple}}
\small
\begin{center}
\begin{tabular}{lcccccc}
\hline
\hline
$z$ &
$\frac{\sigma_{R/s}}{R/s}$ &
$V$ &
$\frac{dN_{ELG}}{dz~ d{\rm deg}^2}$ & $\frac{dN_{LRG}}{dz ~d{\rm deg}^2}$ &
$\frac{dN_{QSO}}{dz~ d{\rm deg}^2}$ & $\frac{\sigma_{f\sigma_8}}{f\sigma_8}$\\
 & \%
 &  ${\rm Gpc}^3$ & & & & \%
 \\
\hline
0.1 & 1.77 & 0.81 & 358 &  38 &   5 & 3.26\\
0.3 & 0.82 & 4.72 & 368 & 126 &  22 & 1.60\\
0.5 & 0.56 & 10.33 & 237 & 333 &  31 & 1.34\\
0.7 & 0.41 & 16.08 & 709 & 570 &  34 & 0.94\\
0.9 & 0.36 & 21.17 & 1436 & 442 &  44 & 0.77\\
1.1 & 0.42 & 25.31 & 1393 &  13 &  56 & 0.80\\
1.3 & 0.41 & 28.51 & 1444 &   0 &  69 & 0.81\\
1.5 & 0.51 & 30.87 & 802 &   0 &  81 & 1.04\\
1.7 & 1.10 & 32.52 & 152 &   0 &  80 & 2.36\\
\hline
\hline
\end{tabular}
\end{center}
\end{table}

\section{Conclusion}

DESI fits perfectly within the established Dark Energy program.  As a Stage-IV BAO experiment, it complements DES and LSST, whose strengths are weak lensing and supernovae.  Building on the solid foundation provided by BOSS, by using a larger telescope and a spectrograph with more fibers and higher resolution, DESI will improve on the Stage III BOSS experiment by an order of magnitude in survey volume.  Although DESI is modest in scale, comparable to the recently completed DECam project, it will be the first project to achieve Stage IV dark energy constraints. DESI comes at the right time; without it there would be a hiatus in dark energy measurement in the U.S. program around 2018--2021. With DESI, the US can maintain its excellence in cosmology, a field that has been transformed by the DOE-supported research of two recent Nobel-Prize efforts.

\end{document}